\documentstyle[12pt,psfig]{l-aa}
\input tcilatex
\newcommand{\beq}{\begin{equation}}
\newcommand{\eeq}{\end{equation}}
\newcommand{\ba}{\begin{array}}
\newcommand{\ea}{\end{array}}

\begin{document}
\pretolerance=5000

\thesaurus{}
\title{Closed and open FRW cosmologies with matter
creation: Kinematic tests}
\author{J. S. Alcaniz and J. A. S. Lima}

\institute{Departamento de Fisica, UFRN, C.P 1641\\
           59072-970 Natal, Brasil }

\date{Received ; accepted}

\offprints{J. A. S. Lima}

\maketitle

\begin{abstract}
The limits imposed by the classical cosmological
tests on closed and open FRW universes driven by adiabatic matter
creation are investigated. Exact expressions for
the lookback time, age of the universe,
luminosity distance, angular diameter,
and galaxy number counts versus redshift are derived
and their meaning discussed in detail. An interesting consequence of
these cosmological models  is the possibility of an accelerated expansion 
today (as
indicated from supernovae observations) with
no need to invoke either a cosmological constant or an exotic 
``quintessence" component.
\end{abstract}

\section{Introduction}

\hspace{0.5cm} It is widely known that the standard cold dark matter FRW 
cosmology present serious theoretical and observational difficulties to be 
considered an acceptable description of the Universe. An overlook in the 
literature shows the existence of a growing body of work discussing 
alternative cosmologies (Carvalho et al. 1992, Krauss and Turner 1995, Lima 
et al. 1996, Caldwell et al. 1997, Overduin and Copperstock 1998).
The first motivation cames from the conflict between the
age of the universe (which is proportional to the Hubble parameter), and 
the age of the oldest stars in globular clusters. The ages of the globular 
clusters tipically fall upon the interval $t_{gc} = (12-14) \rm{Gyr}$ 
(Bolton and Hogan 1995, Pont et
al. 1998, Riess et al. 1999), while measurements of the expansion time 
scale of the Universe
are now converging to $h = (H_{o}/100 \rm{km/sec/Mpc}) = 0.7 \pm 0.1$ 
(Freedman
1998). For this value of the ``little" $h$, the theoretically favoured 
Einstein-de Sitter Universe predicts an age of the Universe ($t_o = {2 
\over3}H_o^{-1})$ within the interval  $8.1 \rm{Gyr} \leq  t_o \leq 10.8
\rm{Gyr}$. For a generic FRW cosmology, the generality of this problem 
comes from the fact that $t_o$ is always smaller than $H_o^{-1}$. Indeed, 
the ``age conflict" is even more acute if we consider its variant based on 
the age constraints from old galaxies at high redshift (Dunlop 1996, Krauss 
1997). As recently argued, the overall tendency is that if more and more 
old redshift galaxies are discovered, the relevant statistical studies in
connection with the ``age problem" may provide very restrictive constraints
for any  realistic cosmological model (Alcaniz and Lima 1999).

Another important piece of data is provided by the recent measurements of 
the
deceleration parameter from SNe Ia observations. Using approximately fifty 
type Ia supernovae, with redshifts between 0 and 1, two groups have 
presented strong evidence that the
universe may be accelerating today ($q_{o} < 0$). This result is in 
apparent contradiction
with a universe filled only by nonrelativistic matter, in such a way that
even open models or more generally, any model with positive deceleration
parameter seems to be in desagreement with these data (Perlmutter et al. 
1998,
Riess et al. 1999).

These problems inspired several cosmologists to consider models with a 
second relic component (an exotic kind of matter, probably of nonbaryonic 
origin) which is seen only by its gravitational effects (Krauss and
Turner 1995, Turner and Write 1997, Chiba et al. 1997). Among these 
scenarios, considerable attention has been dedicated to models
with a  cosmological constant $\Lambda$, a primeval scalar field (Ratra and 
Peebles 1988), decaying vacuum cosmologies (Overduin and Copperstock 1998), 
as well as a noninteracting $x$-component (Silveira and Waga 1997, Caldwell 
et al. 1998).

On the other hand, scenarios with a different kind of ingredient, namely, 
an adiabatic matter creation process, has also been proposed in the
literature (Lima et al. 1996, Lima and Abramo 1999). The limits imposed by 
the classical cosmological tests on a class of dust filled flat FRW 
cosmologies with matter creation have also been examined (Lima and Alcaniz 
1999, hereafter paper I). In this sort of cosmology, the age of the 
universe may be large enough to agree with the observations, and more 
important still, there is no need to invoke  a second smooth component in 
order to generate a negative deceleration parameter.

In this context, the aim of the present work is to extend the treatment of 
the paper I to include both the elliptic ($k=+1$) and hyperbolic ($k=-1$) 
Universes. The paper is organized as follows. Next section we set up the 
basic equations for FRW type cosmologies endowed with an adiabatic matter 
creation process. The classical cosmological tests are described and 
compared with the flat case in section 3.

\section{Universes with Adiabatic Creation : Basic Equations}

We start with the homogeneous and isotropic FRW line element
$(c=1)$
\begin{equation}
 ds^2 = dt^2 - R^{2}(t) (\frac{dr^2}{1-k r^2} + r^2 d \theta^2
+ r^2 sin^{2}\theta d \phi^2) \quad ,
\end{equation}
where $r$, $\theta$, and $\phi$ are dimensionless comoving
coordinates, $k=0$, $\pm 1$ is the curvature parameter of the
spatial sections and $R(t)$ is the scale factor.

In models with ``adiabatic" creation, the dynamic behavior is determined by 
the Einstein field equations (EFE) together the balance equation for the 
particle number density (Prigogine et al. 1989, Calv\~{a}o et al. 1992, 
Lima and Germano 1992)
\begin{equation}\label{rho}
8\pi G \rho = 3 \frac{\dot{R}^2}{R^2} + 3 \frac{k}{R^2} \quad,
\end{equation}
\begin{equation}\label{press}
8\pi G (p+p_{c}) = -2 \frac{\ddot{R}}{R} -
\frac{\dot{R}^2}{R^2}
- \frac{k}{R^2}\quad ,
\end{equation}
\begin{equation}\label{dotn}
 \frac{\dot{n}}{n} + 3 \frac{\dot{R}}{R} =
           \frac{\psi}{n}\quad ,
\end{equation}
where an overdot means time derivative and $\rho$, $p$, $n$ and $\psi$ are 
the energy density, thermostatic pressure, particle number density and 
matter creation rate, respectively. The creation pressure
$p_{c}$ depends on the matter creation rate, and
for ``adiabatic'' creation, it takes the
following form (Calv\~ao et al. 1992, Lima and Germano 1992)
\beq\label{cpress}
p_{c} = - {\rho + p \over 3nH} \psi \quad,
\eeq
where $H = {\dot {R}}/R$ is the Hubble parameter.

To give a complete description, the set (2-5) must be supplemented by an 
equation of state, which in the cosmological domain is usually given by
\begin{equation}
p=(\gamma -1)\rho
\end{equation}
where the $\gamma$ parameter specifies  if
the universe is  radiation ($\gamma={4 \over
3}$) or dust ($\gamma=1$) dominated.

Now, according to paper I (see also Lima et al. 1996), we assume that the 
matter creation rate is
\begin{equation}\label{crate}
\psi = 3 \beta n H \quad ,
\end{equation}
where the $\beta$ parameter  must be
determined either from a kinetic theoretical approach or from quantum field 
theory in
curved spacetime. In general, the $\beta$ parameter is a function of the 
cosmic era, or
equivalently, of the $\gamma$ parameter. Assuming that only the creation of 
the dominant component contributes appreciably to the matter content, we 
should have at least two parameters, $\beta_r$ and
$\beta_m$, for each phase of the Universe (radiation and matter). However, 
since we are particularly interested in the present matter dominated phase, 
from now on we take $\gamma=1$ and $\beta_m=\beta$ supposed to be constant 
and defined on the interval $[0,1]$.

Combining equations (\ref{rho}), (\ref{press}), (\ref{dotn}), {\ref{cpress} 
and
(\ref{crate}), is readily seen that the evolution
equation for the scale factor reads
\begin{equation}
R \ddot{R} + \Delta  \dot{R}^2 + \Delta k = 0\quad ,
\end{equation}
the first integral of which is
\begin{equation}
{\dot{R}}^2 =  \frac{A}{R^{2\Delta}} - k\quad ,
\end{equation}
where $\Delta = \frac{3(1-\beta)-2}{2}$. By expressing the constant $A$ in 
terms of the present day parameters (see Eq.(2)), it is straightforward to 
show that the
above equation can be written as
\begin{equation}
({\dot{R} \over R_{o}})^{2} = H_{o}^{2} \left[1 - \Omega_{o} + 
\Omega_{o}({R_{o} \over
R})^{1 - 3\beta} \right]\quad .
\end{equation}
where $\Omega_o = {\rho \over \rho_c}\vert_{t=t_o}$ and $H_o={\dot R \over 
R}\vert_{t=t_o}$
are the present values of the density and Hubble parameters.
For $\beta=0$ the above equation reproduces the standard cold dark matter 
FRW result (Kolb and Turner 1990).
In virtue of the matter creation, we also see that
the  explicit dependence of the energy density on the scale
factor $R(t)$ is slightly modified in comparison with the standard case. 
Combining (2) and (9) one finds
\begin{equation}\label{rhoR}
\rho = \rho_o
{(\frac{R_o}{R} ) }^{3(1-\beta)}\quad ,
\end{equation}
where $\rho_o = 3A/8\pi GR_{o}^{3(1 - \beta)}$. It thus follows from the 
definition of $\Delta$, that all the expressions of physical interest are 
obtained from the standard ones simply by replacing the ``index" $\gamma 
=1$ by an effective parameter
$\gamma_{eff} = 1 - \beta$. This explains why a dust dominated Universe may 
have a dynamic behavior equivalent to a Universe filled with a matter 
component with negative pressure.

Following standard lines we also define
the deceleration parameter $q_o = - {{R\ddot R \over {\dot 
R}^2}}\vert_{t=t_o}$. Using equations (2), (6) and (7)  one may show that
\beq
q_o = {{1-3\beta}\over 2}\Omega_o \quad.
\eeq
Therefore, for any value of $\Omega_o \neq 0$, we see that the 
decceleration parameter $q_o$ with matter creation is always smaller than 
the corresponding one of the
FRW model. The critical case ($\beta={1 \over 3}, q_o = 0$), describes a 
``coasting cosmology". However, instead of being supported by ``K-matter" 
(Kolb 1989), this kind of model is obtained in the present context for a 
dust filled universe, and the corresponding solutions hold regardless of 
the value
of $\Omega_o$. It is also interesting that even negative values of $q_o$ 
are allowed for a dust filled Universe, since the constraint $q_o < 0$ can 
always be satisfied provided $\beta > 1/3$. These results are in line with 
recent measurements of the deceleration parameter $q_o$ using Type Ia 
supernovae (Perlmutter et al. 1998; Riess et al. 1999). Such observations 
indicate that the universe may be accelerating today, which corresponds 
dinamically to a negative pressure term in the EFE. For a fixed $\Omega_o$, 
this means that the universe with creation is older than the corresponding 
FRW model with the usual decceleration parameter $q_o \geq 0$. This 
behavior also reconcile other recent
results (Freedman 1998), pointing to a Hubble parameter $H_o$ larger than 
$50$ km s$^{-1}$ Mpc$^{-1}$. To date, only scalar field models (Ratra and 
Peebles 1988), and the so-called ``quintessence" (of which $\Lambda$ is a 
special case) have been invoked as being capable of explaining these 
results (Caldwell et al. 1998). As remarked before, in the present 
framework, is the creation pressure that provides the additional 
acceleration measured by a negative $q_{o}$, and not an exotic equation of 
state as in models dominated either by the cosmological constant or a 
``quintessence" ($\gamma < 1$) (Caldwell et al. 1998, Huey et al. 1998).

\section{Kinematic Tests}

The kinematical relation distances must be confronted with the observations 
in order to put limits on the free parameters of the model. Now, we derive 
the kinematical relations for the closed and open cases of the model 
considered here. The limits imposed
by these tests will be compared with the ones imposed on the flat case
(paper I) in the last section.

\begin{figure}
\vspace{.2in}
\centerline{\psfig{figure=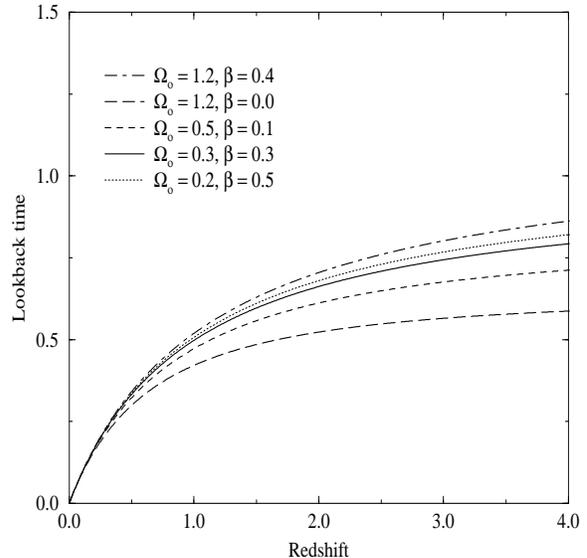,width=3truein,height=3truein}
\hskip 0.1in}
\caption{Lookback time as a function of the redshift for some selected
values of
$\Omega_{o}$ and $\beta$. The lookback time increases for higher values of
$\beta$, i.e., models with
larger matter creation rate are older.}
\end{figure}
\begin{figure}
\vspace{.2in}
\centerline{\psfig{figure=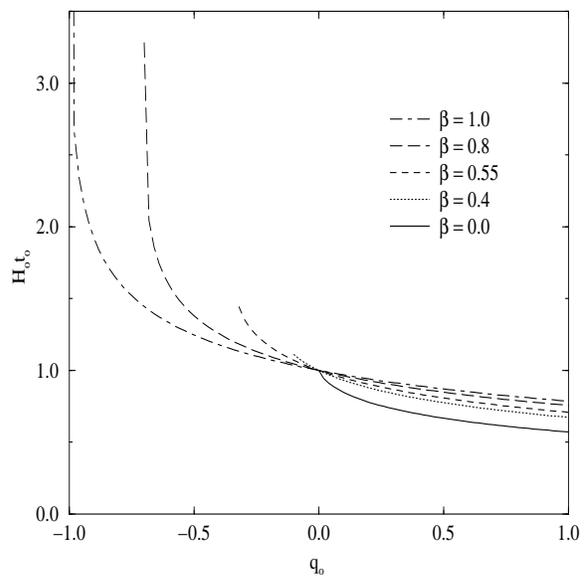,width=3truein,height=3truein}
\hskip 0.1in}
\caption{Age parameter as a function of the deceleration parameter for
some selected values of
$\beta$. Solid curve is the standard FRW Universe with no matter creation
($\beta = 0, q_{o} \geq 0$). It follows from eq.(12) that for $\beta \geq 
1/3$, the deceleration parameter assume negative values.}
\end{figure}

a) Lookback time-redshift Diagram

The lookback time, $\Delta t = t_o - t(z)$, is the difference between the 
age of the Universe at the present time ($z=0$) and the age of the Universe 
when a particular light ray at redshift $z$ was emmited. By integrating 
(10) such a quantity is easily derived
{\small{
\begin{equation}
t_o - t(z) = {H_o}^{-1} \int_{(1+z)^{-1}}^{1}
{[1-\Omega_o + \Omega_o x^{-(1-3\beta)}] ^{-1/2}
dx}
\end{equation}}}
which generalizes the standard FRW result (Kolb and Turner 1989).
The age of the Universe is
obtained by taking the limit
$z \rightarrow \infty$  in  the above equation. We find

\begin{equation}
t_o  = {H_o}^{-1} \int_{0}^{1}
     {[1-\Omega_o + \Omega_o x^{-(1-3\beta)}]^{-1/2} dx}\quad .
\end{equation}

For $\Omega_o=1$ these expressions reduce to the flat case studied in the 
paper I. Generically, we see that matter creation increases the
dimensionless parameter $H_{o}t_{o}$ while preserving the
overall expanding FRW behavior. The lookback time curves as a function of 
the redshift for some selected values of $\Omega_{o}$ and $\beta$
are displayed in Fig. 1. For completeness, in Fig. 2 we show the age of the 
Universe (in units of $H_{o}$) as a function of the deceleration parameter.

b) Luminosity distance-redshift

The luminosity distance of a light source is defined as the ratio of the
detected energy flux $L$, and the apparent luminosity, i.e., $d_l^{2} = {L
\over 4 \pi l}$. In the standard FRW metric (1) it takes the form
(Sandage 1988)
\begin{equation}
d_l = R_o r_1(z)(1 + z)\quad ,
\end{equation}
where $r_1(z)$ is the radial coordinate distance of the object
at light emission. Inserting $r_1(z)$ derived in the Appendix, it
follows that
\begin{equation}
 d_l = \frac{(1+z)sin[\delta sin^{-1}(\alpha_1)-\delta
sin^{-1}(\alpha_2)]}{H_o(\Omega_o-1)^{\frac{1}{2}}}\quad .
\end{equation}
where $\delta=\frac{2}{(1-3\beta)}$, $\alpha_1=
(\frac{\Omega_o-1}{\Omega_o})^{{1 \over 2}}$,
and $\alpha_{2}=\alpha_{2} (1 + z)^{-\frac{1-3\beta}{2}}$.

As one may check, expressing $\Omega_o$ in terms of $q_o$ from
(10), and taking the limit $\beta \rightarrow0$, the above
expression reduces to
\begin{equation}
 d_l = \frac{1}{{H_o}q_o^{2}}[zq_o + (q_o - 1)(\sqrt{2q_oz + 1}
-
1)]\quad ,
\end{equation}
which is the usual FRW result (Weinberg 1972). Expanding eq.(14) for small 
$z$
gives
\begin{equation}
H_o d_l = z + \frac{1}{2}(1 - \frac{{1 - 3\beta}}{2}\Omega_o)
z^{2} +...\quad ,
\end{equation}
which depends explicitly on the matter creation $\beta$
parameter. However, replacing $\Omega_o$ from (10) we recover
the usual FRW expansion for small redshifts, which depends only on
the effective deceleration parameter $q_o$ (Weinberg 1972). This is
not a surprizing result since expanding $d_l(z)$ in terms of $\Omega_o$, 
the OO component of
Einstein's equations has implicitly been considered, while the expansion in 
terms of
$q_o$ comes only from the form of the FRW line element. The
luminosity distance as a function of the redshift for closed and open 
models
with adiabatic matter creation is shown in Figures 3 and 4, respectively.
As espected for all kinematic tests, different cosmological models have
similar behavior at $z << 1$, and the greatest discrimination among them 
comes
from observations at large redshifts.

\begin{figure}
\vspace{.2in}
\centerline{\psfig{figure=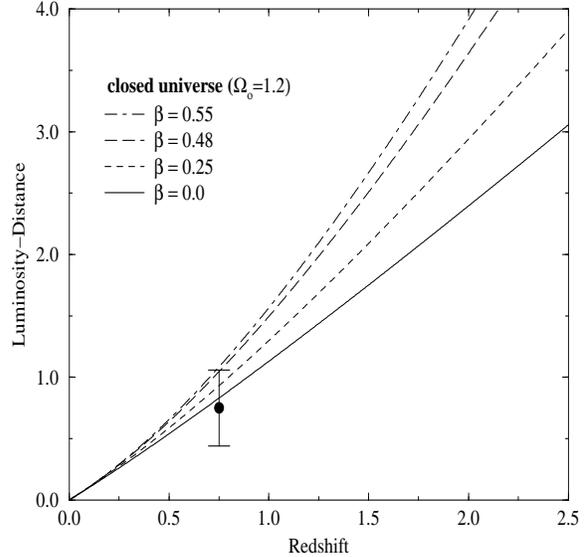,width=3truein,height=3truein}
\hskip 0.1in}
\caption{Luminosity distance as a function of the redshift for closed
models with adiabatic
matter creation. Solid curve is the standard FRW closed model
($\beta = 0$). The
selected values of $\beta$ are shown in the picture. Here and in Fig.
4 the typical error
bar and data point are taken from Kristian et al. (1978).}
\end{figure}
\begin{figure}
\vspace{.2in}
\centerline{\psfig{figure=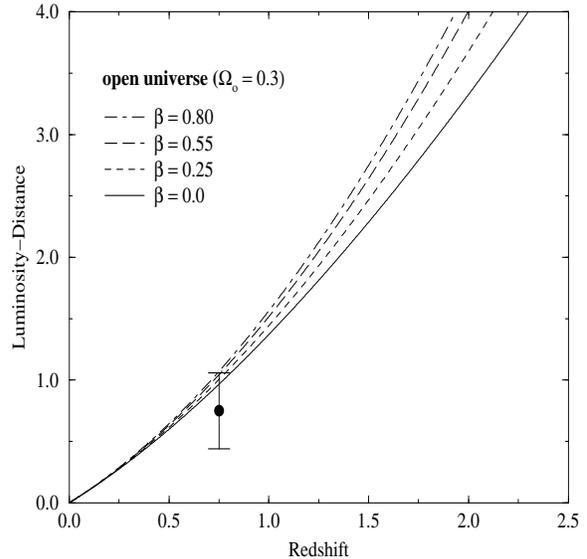,width=3truein,height=3truein}
\hskip 0.1in}
\caption{The same graph of Fig. 3 for open models. Solid curve is
the standard FRW open
model ($\beta = 0$).}
\end{figure}

c) Angular size-redshift

Another important kinematic test is the angular size - redshift relation
($\theta(z)$). As widely known, the data concerning the angular-size are 
until nowadays
somewhat controversial (see Buchalter et al. 1998 and references therein).
Here we are interested in  angular diameters of light sources described as
rigid rods and not as isophotal diameters. These quantities are naturally
different,  because in an expanding world the surface brightness varies
with the distance (Sandage 1988).
The angular size of a light source of proper size $D$ (assumed free of 
evolutionary effects)  located at $r = r_1(z)$ and observed at $r = 0$ is
\begin{equation}
\theta = \frac{D(1 + z)}{R_o r_1(z)}\quad .
\end{equation}
Inserting  the expression of $r_1(z)$ given in the Appendix it follows that
{\small{
\begin{equation}
\theta = {DH_o(\Omega_o - 1)^{\frac{1}{2}}(1 + z)}{{sin[\delta
sin^{-1}\alpha_2-\delta sin^{-1}\alpha_1]}}\quad .
\end{equation}
}}

For small $z$ one finds
\begin{equation}
\theta = {D H_o \over z}[ 1 + \frac{1}{2}
(3 + \frac{{1 - 3\beta}}{2}\Omega_o) z +...]\quad .
\end{equation}
Hence, ``adiabatic" matter creation as modelled here also
requires an angular size decreasing as the inverse of the
redshift for small $z$. However, for a given value of
$\Omega_o$,
the second order term is a function only of the $\beta$ parameter.
In terms of $q_o$, inserting (10)
into (19) it is readily obtained
\begin{equation}
\theta = \frac{D H_o}{z}[ 1 + \frac{1}{2}
(3 + q_o) z +...]\quad ,
\end{equation}
which is formally the same FRW result for small redshifts
(Sandage 1988). At this limit only the effective deceleration
parameter may be constrained from the data, or equivalently, at
small redshifts one cannot extract the values of $\Omega_o$ and
$\beta$ separately. The angular size-redshift diagram for closed and open
models and selected values of the $\beta$ parameter is displayed in Figures 
5
and 6, respectively.

\begin{figure}
\vspace{.2in}
\centerline{\psfig{figure=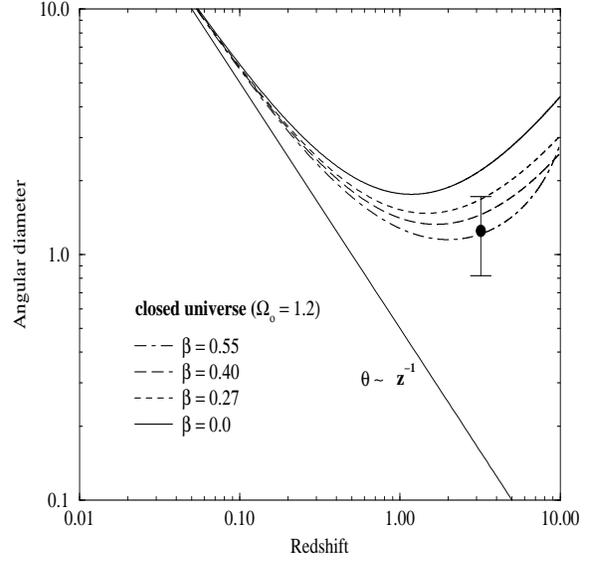,width=3truein,height=3truein}
\hskip 0.1in}
\caption{Angular diameter versus redshift in closed models with
adiabatic matter creation
and some selected values of $\beta$. Solid curve is the standard
model ($\beta = 0$).
The angular size reaches a minimum at a given $z_{c}$ and increases
for fainter magnitudes.
The minimum is displaced for higher $z$ as the $\beta$ parameter is
increased. The straight line is the Euclidian result. Here and in
Fig. 6, the typical error bar and data point are taken from Gurvits
(1994).}
\end{figure}

\begin{figure}
\vspace{.2in}
\centerline{\psfig{figure=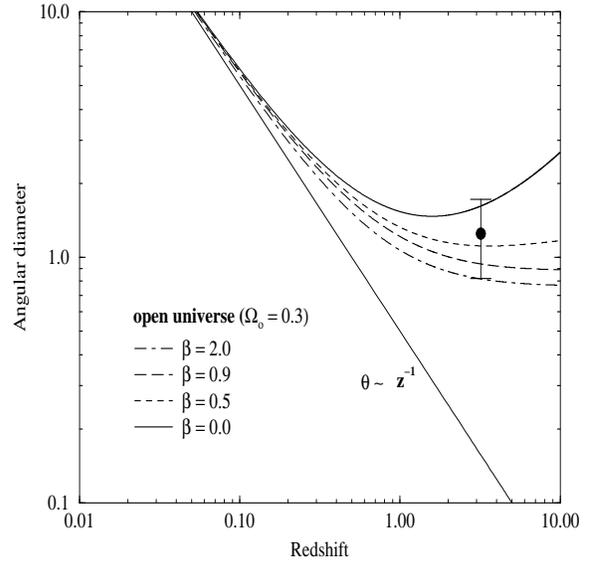,width=3truein,height=3truein}
\hskip 0.1in}
\caption{Angular diameter versus redshift: the open case. As in Fig.5, the 
straight line is the Euclidian result. Solid curve is the prediction of the 
standard FRW open universe ($\beta = 0$).}
\end{figure}

d) Number counts-redshift

The final kinematic test considered here is the galaxy number count per
redshift interval. We first notice that
although modifying the evolution equation driving the
amplification of small perturbations, and so the usual
adiabatic treatment for galaxy formation, the created matter is smeared
out and does not change the total number of sources present in the
nonlinear regime. In other words, the number of galaxies
already formed scales with $R^{-3}$ (Lima et al. 1996, Lima and Alcaniz 
1999).

The number of galaxies in a comoving volume is equal to the number density 
of
galaxies (per comoving volume) $n_{g}$, times the comoving volume element
$dV_{c}$
\begin{equation}
dN_{g}(z) = n_{g} dV_{c} = \frac{n_{g} r^{2} dr d{\Omega}
}{\sqrt{1 - k{r}^{2}}}\quad .
\end{equation}

By using $r_1(z)$ as derived in appendix, it follows that the general
expression for number-counts can be
written as
\begin{equation}
{(H_oR_o)^{3}dN_{g}\over n_{g}z^{2}dzd{\Omega}} =
{sin^{2}{\delta[sin^{-1}
(\alpha_2-sin^{-1}(\alpha_1)]}
\over (1 + z)z^{2}f(\Omega_o, \beta, z)}\quad ,
\end{equation}
where $f(\Omega_o, \beta, z) = (\Omega_o - 1)[1-\Omega_{o} +
\Omega_o(1+z)^{1-3\beta}]^{1/2}$.
For small redshifts
\begin{equation}
\frac{(H_oR_o)^{3} dN_{g}}{n_{g}z^{2}dzd{\Omega}} =
1 - 2(\frac{\Omega_o(1-3\beta)}{2} + 1)z + ...\quad .
\end{equation}
In Figures 7 and 8, we have displayed the number counts-redshift relations
of closed and open Universes for some selected values of $\Omega_{o}$ and 
$\beta$. It is worth
mentioning the tendency of matter creation models to have larger volumes
per redshift interval than the standard FRW models with the same
$\Omega_{o}$. This feature is similar to the one found in decaying vacuum 
cosmologies and
could turn out to be advantageous if the
observational data indicate an excess count of high-redshift objects (Waga 
1993). The limits
on the $\beta$ parameter obtained from all
kinematic tests are shown in Table 1.

\section{Conclusion}

The recent observational evidences for an accelerated state of the present
Universe, obtained from distant SNe Ia (Perlmutter et al. 1998) give a 
strong support to the search of alternative cosmologies. As demonstrated 
here, the process of adiabatic matter creation is also an ingredient 
accounting for this unexpected observational result. In a previous analysis 
(Lima and Alcaniz 1999) we have examined such a possibility for a flat 
Universe, while in the present paper we extend all the analysis for closed 
and open cosmologies. In this way, the expanding ``postulate'' and its main 
consequences may also be compatibilized with a cosmic fluid endowed with 
adiabatic matter creation.

The rather slight changes introduced by the matter creation process, which 
is quantified by the $\beta$ parameter, provides a reasonable fit of 
several cosmological data. Kinematic tests like luminosity distance, 
angular diameter and number-counts versus redshift relations constrain 
perceptively the matter creation parameter. For models charactherized by 
the pair ($\Omega_o, \beta$), the age of the Universe is always greather 
than the corresponding FRW model ($\beta=0$), and even values bigger than 
$H_o^{-1}$ are allowed for all values of the curvature parameter. However, 
in spite of these important physical
consequences, the matter creation rate nowadays, $\psi_o =
3n_oH_o \approx 10^{-16}$ nucleons $cm^{-3}yr^{-1}$, is nearly
the same rate predicted by the steady-state Universe (Hoyle et
al. 1993) regardless the value of the curvature parameter. This matter 
creation rate is presently far below detectable limits.

\begin{figure}
\vspace{.2in}
\centerline{\psfig{figure=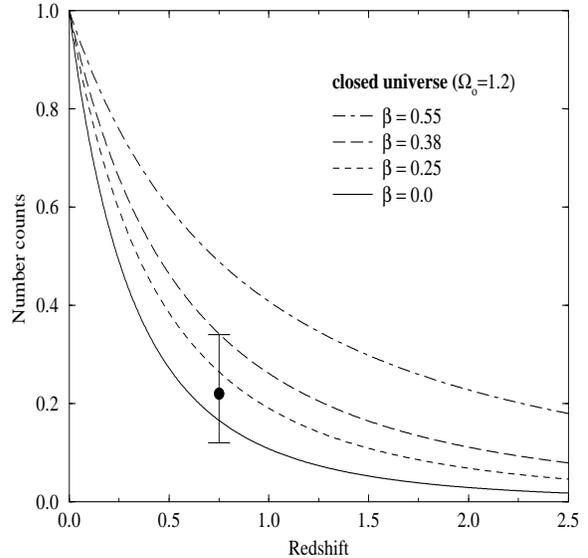,width=3truein,height=3truein}
\hskip 0.1in}
\caption{Number counts as a function of the redshift for closed models
with adiabatic matter
creation. All results are shown for $\Omega _{o} = 1.5$ and some
values of $\beta$. Solid curve is the closed FRW model with no matter 
creation. Here and
in Fig. 8, the typical
error bar and data point are taken from Loh and Spillar (1986).}
\end{figure}
\begin{figure}
\vspace{.2in}
\centerline{\psfig{figure=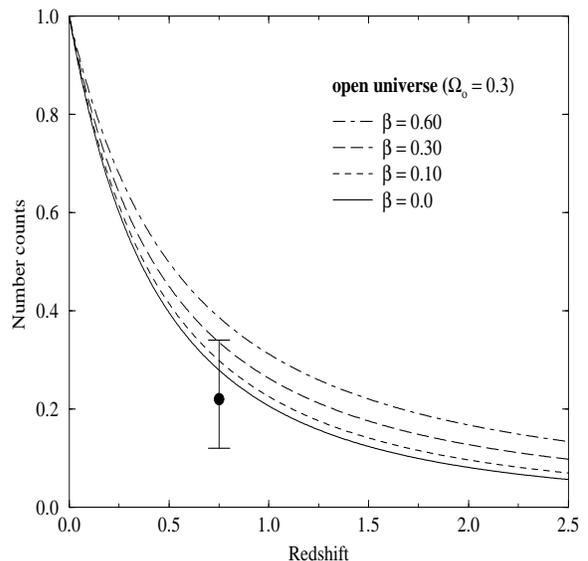,width=3truein,height=3truein}
\hskip 0.1in}
\caption{Number counts versus redshift: the open case. As in Fig.7, solid 
curve is the prediction of the standard FRW open universe ($\beta = 0$).}
\end{figure}
\begin{table}[t]
\caption{Limits to $\beta$}
\begin{tabular}{rll}
\hline
\\
\multicolumn{1}{c}{Test}&
\multicolumn{1}{c}{Open}&
\multicolumn{1}{c}{Closed}\\
\\
\hline
\\
Luminosity distance-redshift&
$\beta \le 0.55$& $\beta \le 0.48$\\
\\
Angular size-redshift&
$\beta \le 3.0$& $\beta \ge 0.27$\\
\\
Number counts-redshift&
$\beta \le 0.30$& $\beta \le 0.38$\\
\\
\hline
\end{tabular}
\end{table}

\appendix
\section{Dimensionless radial coordinate versus redshift relation}

Some observable quantities in the standard FRW model are
easily determined expressing the radial dimensionless
coordinate
$r$ of a source light as a function of the redshift (Mattig
1958).
In this appendix, we derive a similar equation to the matter
creation scenario discussed in this paper.

Now consider a typical galaxy located at
$(r_1,\theta_1,\phi_1)$
emitting radiation to an abserver at $(0,\theta_1,\phi_1)$. If
the waves leave the source at time  $t_1$ and reach the
observer
at time $t_0$, the null geodesic equation ($dt^2 - \frac{R^{2}dr^2}{1-k 
r^2} =
0$),
which define the light track yields
\begin{equation}
\int_{t_0}^{t_1}
     \frac{dt}{R(t)} = \int_{0}^{r_1}
     \frac{dr}{\sqrt{1 - k{r}^{2}}}= \frac{Arcsin\sqrt
kr_1}{\sqrt k} = I\quad .
\end{equation}
Since $t=t(R)$, changing variable to $x={R \over R_o}$ and
using (10), the above result reads
\begin{equation}
I = \frac{1}{R_o H_o}
\int_{(1+z)^{-1}}^{1}
           { [ 1-\Omega_o + \Omega_o x^{-(1-3\beta)} ] ^{-1/2}
{dx \over x} }\quad .
\end{equation}

This integral depends on the values of the $\Omega_o$ and
$\beta$ parameters. For $\beta = \frac{1}{3}$ one finds the same
results of the coasting cosmology (Kolb 1989). For $\beta$ different of
${1 \over 3}$, we introduce a new auxiliar variable $y^{2} =
(\frac{\Omega_o - 1}{\Omega_o})x^{(1-3\beta)}$, in terms of
which the above equation becomes
\begin{equation}
\frac{Arcsin\sqrt kr_1}{\sqrt k} = \frac{\delta}{R_o
H_o(\Omega_o
- 1)^{\frac{1}{2}}}\int_{\alpha_1}^{\alpha_2}
           \frac{dy}{\sqrt{1 - {y}^{2}}}\quad ,
\end{equation}
where $\delta = \frac{2}{(1-3\beta)}$, $\alpha_{2} =
(\frac{\Omega_o-1}{\Omega_o})^{\frac{1}{2}}$, and
$\alpha_1 = \alpha_{2}(1+z)^{-\frac{(1-3\beta)}{2}}$.

The right hand side of the above integral is the same appearing
in (A1) for $k=1$. Hence, replacing in (A3) the value of $k$ given by (2) 
and (9), it is readily seen that
\begin{equation}
r_1(z) = \frac{sin[\delta sin^{-1}\alpha_2-
\delta sin^{-1}\alpha_1]}{R_oH_o(\Omega_o -
1)^{\frac{1}{2}}}\quad .
\end{equation}
In particular, the limit for a flat Universe ($\Omega_o=1$)
yields (paper I)
\begin{equation}
r_1(z)= {\frac{2}{(1-3\beta)R_oH_o}\{1 - (1 + z)^{\frac{2}{{1
-3\beta}}}\}}\quad ,
\end{equation}
which could have been obtained directly from (A2).

In terms of the deceleration parameter (A4) may be rewritten
as
\begin{equation}
r_1(z) = \frac{sin[\delta sin^{-1}\alpha_2-
\delta sin^{-1}\alpha_1]}{R_oH_o(\frac{2q_o}{1 - 3\beta} -
1)^{\frac{1}{2}}}\quad ,
\end{equation}
which in the limit $\beta \rightarrow 0$ reduces to the usual FRW result
(Weinberg 1972)
\begin{equation}
r_1(z) = \frac{q_oz + (q_o - 1)(\sqrt{2q_oz + 1} -
1)}{H_oR_oq_o^{2}(1 + z)}\quad .
\end{equation}
Equation (A4), or equivalently (A6), plays a key role in the
derivation of some astrophysical quantities discussed in this
paper.

\begin{acknowledgements}
This work was partially
supported by the project Pronex/FINEP (No. 41.96.0908.00) and
Conselho Nacional de Desenvolvimento Cient\'{\i}fico e
Tecnol\'ogico - CNPq (Brazilian Research Agency).
\end{acknowledgements}

\end{document}